# Observation of Electronic Modes in Open Cavity Resonator

Hwanchul Jung[1*], Dongsung T. Park[2*], Seokyeong Lee[2], Uhjin Kim[3], Chanuk Yang[3], Jehyun Kim[4], V. Umansky[5], Dohun Kim[4], H.-S. Sim[2], Yunchul Chung[1†], Hyoungsoon Choi[2‡], Hyung Kook Choi[3§]

[1]Department of Physics, Pusan National University, Busan 46241, Republic of Korea.
[2] Department of Physics, KAIST, Deajeon 34141, Republic of Korea.
[3] Department of Physics, Research Institute of Physics and Chemistry, Jeonbuk National University, Jeonju 54896, Republic of Korea.
[4] Department of Physics and Astronomy, Seoul National University, Seoul 08826, Korea.
[5] Department of Condensed Matter Physics, Weizmann Institute of Science, Rehovot 76100, Israel.

The resemblance between electrons and optical waves has strongly driven the advancement of mesoscopic physics. However, electron waves have yet to be understood in open cavity structures which have provided contemporary optics with rich insight towards non-Hermitian systems and complex interactions between resonance mode. Here, we report the realization of an open cavity resonator in a two-dimensional electronic system. We studied the resonant electron modes within the cavity and resolved the signatures of longitudinal and transverse quantization, showing that the modes are robust despite the openness of the cavity being highly open to the background continuum. The transverse modes were investigated by applying a controlled deformation to the cavity, and their spatial distributions were further analyzed using magnetoconductance measurements and numerical simulation. These results lay the groundwork to exploring electronic wavefunctions in the context of modern optical systems, such as the dielectric microcavity.

[*] These authors contribute equally.
[†] ycchung@pusan.ac.kr
[‡] h.choi@kaist.ac.kr
[§] hkchoi@jbnu.ac.kr

The profound resemblance between optical waves and ballistic electrons has inspired numerous electronic realizations of optical elements such as lenses[1–3], beam splitters[4–7], and their implementation towards various electron interferometers[8–17]. In particular, the interference between electronic quasiparticles has become an especially important concept when probing various exotic properties of many-body excitations, e.g. anyon braiding in fractional quantum Hall excitations[18–23]. Such interferences arise naturally in optical physics under the name of cavity resonances[24–26], and recent developments in optical microcavities have emphasized the role of cavity modes in understanding integrable ray dynamics and wave behavior in semi-open systems[27–30]. Although mode resonances in cavity-like structures have also been studied in electronic systems, much research had only focused on nearly-closed structures, e.g. quantum dots (QDs) and quantum billiards[31–33]. Even in the celebrated quantum Hall Fabry-Pérot configuration[12,18], the edge states serve a purpose similar to fiber optic cables where transverse motion is strongly restricted. However, the richness of mode dynamics in optical cavities arise from the availability of multiple resonator dimensions and the deliberate openness of the cavities[28–30]. While optical modes are well-understood, their adaptation to an electronic open cavity is a nontrivial problem that has yet to be addressed.

Here, we report the realization of the open cavity resonator in a GaAs/AlGaAs two-dimensional electron gas (2DEG) using curved split-gates to define the cavity mirrors. Electrons were injected to the cavity using tunnel-coupled quantum point contacts (QPCs) at the center of the mirrors, and a modulation gate covering the cavity region was used to control the electron wavelength. Cavity resonances were observed from conductance measurements after verifying the strong coupling of the cavity to its open sides. From the conductance lineshape analysis and resonance energy spectrum, we characterized the longitudinal modes as Fabry-Pérot resonances. Furthermore, we demonstrated the tunability of transverse modes by individually controlling the mirror split-gates in order to introduce a cavity deformation. Deforming the cavity induced a detuning among the resonances, and the spatial distribution of these modes were probed by magnetoconductance measurements. With the aid of tight-binding simulations, we identified that the observed transverse modes came in two variants: one lying on top of the central cavity axis, and one lopsided to a side of the axis. Curiously, we found that both modes coexist within the cavity and that the dominant mode undergoes a transition as a function of the electron wavelength.

## Results

**Electronic open cavity resonator.** An optical cavity resonator is also called a Fabry-Pérot resonator since longitudinal cavity modes can be described as Fabry-Pérot interferences along the central cavity axis[25,26]. Although an archetypical Fabry-Pérot interferometer is constructed with two flat mirrors placed parallel to each other (Fig. 1a), such an arrangement is rarely used because small misalignments in the mirrors can easily let the rays escape through the sides and spoil the cavity resonance. Instead, the mirrors can be replaced with mirror lenses that refocus the diverging rays back towards the central cavity axis (Fig. 1b), hence raising the likelihood that the ray stays within the cavity. For mirror radii of $R_1$ and $R_2$, ray optics predicts that the cavity is stable if $0 \leq (1 - L/R_1)(1 - L/R_2) \leq 1$ where $L$ is the cavity length. The lensing action of the cavity mirrors can be thought of as a confinement potential in the transverse direction[34], and rays in the cavity experience an effective confinement in all directions: by reflections in the longitudinal direction and lensing in the transverse direction. In wave optics, these confinements give rise to standing waves, i.e. modes. In particular, the resonance of longitudinal modes is well-recognized as transmission peaks in a Fabry-Pérot spectrum.

Figure 1c is a false-colored image of the open cavity device, fabricated on a GaAs/AlGaAs heterostructure with a 2DEG of density $n = 2.3 \times 10^{11}$ cm$^{-2}$ and mobility $\mu = 3.6 \times 10^6$ cm$^2$/Vs, residing 71 nm under the wafer surface. Metallic Schottky gates (false-colored yellow) were defined on the surface using standard electron beam lithography. Four gates (1u, 1d, 2u, 2d) were negatively biased to locally deplete the 2DEG and form the cavity resonator. Specifically, gates 1u-1d (2u-2d) functioned as the left (right) mirror lens, and the area between the mirrors corresponds to the resonant region within the cavity. The cavity dimensions were $R_{1,2} = 350$ nm and $L = 500$ nm which satisfy the aforementioned cavity stability condition. A tunnel-coupled QPC was formed at the center of the mirrors where the electrons were partially transmitted into or out of the cavity. The Fermi energy in the cavity was controlled by a voltage $V_M$ applied to the modulation gate (M), i.e. $E_F = E_{F,0} + \alpha V_M$ where $\alpha$ is some proportionality factor. The 2DEG reservoir is divided into four parts, the source (S), drain (D), and open sides (O1 and O2), and the cavity properties were measured from the conductance between these reservoirs. The conductance measurements were done using a homemade current preamplifier[35], followed by a lock-in amplifier with an excitation voltage of 10 µV$_{rms}$ at 489 Hz. All measured conductances $G$ are presented in their normalized forms $g = G \times (2e^2/h)^{-1}$ for brevity, and the conductance of the mirror QPC 1u-1d (2u-2d) is denoted as $g_1$ ($g_2$). The experiments were performed in a homemade dilution refrigerator with a base temperature $< 150$ mK.

Figure 1d shows the conductance of the cavity measured while varying $V_M$. Since the modulation gate directly affects the mirror QPCs, the QPC gate voltages were linearly adjusted to compensate for changes in $g_1$ and $g_2$. For example, $V_{1u} = V_{1d} = V_1 + \beta \Delta V_M$ where $\beta$ is an empirical factor keeping $g_1$ relatively constant. We first measured the conductance between O1 and O2, $g_{open}$ (purple line in Fig. 1d), to confirm that the cavity is open in its transverse directions. Even when the cavity was highly open to its sides, $g_{open} \gtrsim 4$ for $V_M \gtrsim 0.15$ V, multiple resonance peaks were observed from the conductance between S and D, $g_{cav}$ (blue line in Fig. 1d), signifying the formation of resonant modes within the cavity. These resonances were further analyzed by varying the QPC conductances: for data presented hereon, the QPC conductances were fixed by nonlinear adjustments to the mirror gate voltages (Supplementary Fig. S2). Figure 2a shows $g_{cav}$ for $g_{1,2} = 0.25 \sim 0.75$ (blue to red lines), and we observed that $g_{cav}$ always decreased with lower $g_{1,2}$. This is rather different from expectation: in a Fabry-Pérot resonator, lowering the mirror transmission generally leads to sharper resonances with increasingly larger transmission peaks[26].

**Longitudinal Fabry-Pérot modes.** We attribute this behavior to the diffraction losses inherent to open cavity resonators[25,26,36]. The electron density in our device corresponds to a Fermi wavelength of 52 nm, but the electrons inside the cavity are expected to have larger Fermi wavelengths due to the combined effect of sample bias cooling and the modulation gate bias. Since the cavity length, 500 nm, is no larger than an order of magnitude from the Fermi wavelength, we expect the diffractive properties of electron waves to be relevant. When such a wave hits the cavity mirror, the reflected wave propagates towards the opposite mirror while spreading in a diffractive manner, and some of the amplitude is unavoidably lost to the open sides. This is in stark contrast to quantum Hall edge Fabry-Pérot interferometers where electrons are either in the 'cavity' or in the source/drain leads. As shown in Fig. 2b, the diffraction can be treated in the Landauer-Büttiker formalism by including a coupling between the cavity modes to its open sides[37], which leads to the transmission amplitude

$$t = t_2 \frac{1}{1 - u_1 r_1 u_2 r_2} u_1 t_1 \qquad (1)$$

where $t_i$ ($r_i$) is the transmission (reflection) amplitude of mirror $i$ and $u_i$ the amplitude acquired while propagating from mirror $i$ to the opposite mirror (Supplementary Fig. S1). In a system with no diffraction loss, the propagation amplitude is simply given by $u_i = \exp(ikL)$ where $k$ is the wavenumber, and the usual lossless Fabry-Pérot interference is recovered. However, the diffraction loss leads to a subunitary propagation, $|u_i|^2 < 1$, and induces a broadening in the transmission peaks. In Fig. 2c, we have plotted the conductance lineshape expected from a cavity with $|u_{1,2}| = 0.66$ and $|r_{1,2}|^2 = 0.25 \sim 0.75$ as red to blue lines. For comparison, we have also plotted the lossless case ($|u_i| = 1$) for $r_{1,2} = 0.75$ as a blue dashed line. The grey points in the figure are the experimental datapoints from the circled peaks in Fig. 2a after normalization. We find that the experimental data is much better described by the lossy Fabry-Pérot model and therefore that the electron wave within the cavity is diffractive. In addition, we highlight another consequence of diffraction within the cavity: the diffraction loss is also proportional to $(1 - u_1 r_1 u_2 r_2)^{-1}$ (see Supplementary Materials for details), and the cavity loss should resemble the transmission spectrum. Indeed, we observed that the conductance from S to O1 and O2, $g_{loss}$ (orange line in Fig. 1c), peaked in concurrence with $g_{cav}$.

Figure 3a shows the source-drain bias spectroscopy obtained by measuring $g_{cav}$ as a function of $V_M$ while applying a voltage bias $V_{SD}$ to the source reservoir (S in Fig. 1c). The open sides (O1, O2 in Fig. 1c) were floated so that the voltage drops occurred only across the QPCs. From the full measurement (inset of Fig. 3a), we have analyzed the part where the conductance peaks were sharply defined with no additional fine structures. The bright regions at $V_{SD} = 0$ indicate the cavity resonances, and a finite $V_{SD}$ splits the conductance peak into negative and positive sloped parts, respectively corresponding to the modes resonating with the biased and grounded Fermi levels. Similar to the Coulomb diamond of QDs[31], the peaks trace a diamond shape (grey lines), and the height of the diamond from $V_{SD} = 0$ gives the energy level spacing $\Delta E_n$ between the neighboring resonances, labeled as $n = 1,2, ...$ in Fig. 3a. In Fig. 3b, we have plotted the level spacings and see that $\Delta E_n$ increases linearly with $n$. Linearity in $\Delta E_n$ implies a quadratic energy spectrum, i.e. $E_n \propto n^2$, which is easily recognizable as the dispersion of free electrons: $E = (\hbar k)^2/2m^*$ where $m^*$ is the effective mass. In a Fabry-Pérot resonator, the electrons acquire a dynamic phase $2kL$ after making a roundtrip between the mirrors, and including the reflection phase $\phi_0$ gives us the resonance condition $2k_n L + \phi_0 = 2\pi n$ and $\Delta E_n = (\hbar^2 \pi^2/m^* L^2) \times (n + n_0 + 1/2)$ where $n_0 = \phi_0/\pi$. The linear fit of the data gives $L = 498$ nm which is in excellent agreement with our cavity length. That is, the observed Fabry-Pérot resonances originate from longitudinal cavity modes.

**Transverse cavity modes.** Having established the longitudinal Fabry-Pérot modes, investigating the transverse modes is the natural next step to understanding electronic cavity modes. Transverse modes are defined by an effective confinement from the mirror lenses and are especially important in understanding cavity dynamics, such as semiclassical quantization and avoided resonance crossings[27–30,36,38–42]. We controlled the transverse modes by introducing a deformation to the mirror geometry. Consider the left mirror gates (1u-1d in Fig. 1a). In an ideal case, the cavity mirror is symmetric when the gates voltages are equal, but placing a more negative voltage on 1d pushes the lower half of the mirror into the cavity. Simultaneously placing a less negative voltage on the 1u pulls the upper half of the mirror away from the cavity, maintaining the average length of the cavity. As illustrated in Fig. 4a, we can parametrize the cavity deformation with the voltage difference between upper and lower mirror gates, e.g. $\delta V_{1(2)} = V_{1(2)u} - V_{1(2)d}$. Figure 4b plots $g_{cav}(V_M)$ measured for various $\delta V = \delta V_1 = \delta V_2$. Near $\delta V \approx -0.1$ V, we see a regular set of Fabry-Pérot resonances as expected. However, the resonance peaks start splitting into two parts as we move up towards a more positive $\delta V$. One set moves leftwards (red arrows), and the other set moves rightwards (blue arrows). This continues until the two set of peaks meet again near $\delta V \approx 0.1$ V. In optical cavities, such peak splitting is commonly understood as transverse mode detuning: each transverse mode corresponds to a set of Fabry-Pérot resonances, and a perturbation to a specific transverse mode applies a phase shift to all the corresponding longitudinal modes. Since the cavity deformation $\delta V$ specifically breaks the reflection symmetry across the central cavity axis, we expect a detuning between modes with different transverse distributions. Interestingly, we found a wavelength-dependent transition in the dominant transverse mode. As a concrete example, we have replotted the data for $\delta V = 0$ V in Fig. 4c. The peak positions have been marked with blue and red circles, and the change in peak heights have been traced with dashed lines. At low $V_M$, i.e. at large Fermi wavelengths, the cavity transmission is initially dominated by the blue set of peaks. Moving towards higher values of $V_M$, however, the blue peaks diminished while the red peaks increased, eventually dominating the cavity conductance.

In order to investigate the spatial distribution of these modes, we measured the magnetoconductance of the cavity device. A Lorentz force was applied to the electrons using an out-of-plane magnetic field $B$ (Fig. 5b). As illustrated in Fig. 5c, a mode lying on the central cavity axis responds (nearly) symmetric to both positive or negative magnetic fields. However, the response of a lopsided mode is expected to distinguish the sign of $B$. Consider an electron launched from the left mirror, guided towards the upper half of the right mirror (Fig. 5d). A positive $B$ would deflect the electron upwards and increase the cavity loss to the open sides. On the other hand, a negative $B$ would guide the electron back towards the mirror center and enhance the conductance peak. That is, the spatial distribution of transverse modes can be known from the sign of $B$ at which $g_{cav}$ shows a maximum. Figure 5a shows the magnetoconductance for a cavity with a positive detuning voltage $\delta V > 0$ similar to the illustrations Figs. 5c, d. As in Fig. 4, we observed two sets of Fabry-Pérot resonances. Inspecting the data at $B = 0$ mT, we have marked the set of peaks dominant at low $V_M$ with blue circles and those dominant at high $V_M$ with red circles. The two sets showed clearly different responses to the magnetic field. The blue-circled peaks were centered at $B \approx 0$ mT while the red-circled peaks were maximized at $B \approx -50$ mT $\sim -20$ mT (red crosses), indicating that the transverse modes of red-circled peaks were located in the upper half of the cavity.

The geometric properties of transverse modes were also reproduced in simulation. Using the numerical tight-binding calculation package KWANT[43], we simulated $g_{cav}$ of an open cavity resonator tuned to similar conditions (Supplementary Fig. S3). In Fig. 5e, we have marked the conductance peaks with blue if the maximum was located at $B = 0$ mT and with red if otherwise. A peak was selected from each set, marked as S for symmetric or A for asymmetric, from which the eigenchannel wavefunction from the left to the right QPCs were obtained. Figure 5f, g shows the calculated wavefunction densities with black areas indicating the mirror gate positions. Indeed, the wavefunction density for peak S (Fig. 5f) lay on the central cavity axis, drawn as a black dashed line, while that of peak A (Fig. 5g), predominantly occupied the upper half of the cavity.

## Discussion

We emphasize that the observed energy spectrum (Fig. 3) markedly distinguishes the open cavity resonator from a QD[31–33]. In a QD, the resonance energy has two components: the charging energy and the quantization energy. The charging energy arises from a mean-field description of the electron-electron interactions and generally decreases with the addition of electrons. On the other hand, the quantization energy is understood in two regimes: the few- and many-electron cases. In the few-electron regime, the quantization energy is usually described in relation to the Fock-Darwin spectrum, which implicitly assumes that the kinetic energy of the electrons is comparable to the potential energy from the confinement. Therefore, the modes of few-electron QDs are ill-suited for an optical analogy which demands that the waves be in free propagation. Moving towards the many-electron regime, the electrons behave more billiard-like as the preoccupying QD electrons provide a shielding from the confinement potential. However, the resonant levels typically become unpredictable as the electron trajectories exhibit chaotic motion, and the geometry of QD modes quickly loses meaning. In an open cavity resonator, however, the charging effect is absent, and the formation of longitudinal modes by free propagating electrons is clearly identified by $\Delta E_n \sim n$. These properties distinguish the electronic cavity device as the preferable platform to study electron waves in the context of quantum billiards and cavity modes.

Returning to the cavity mode distribution (Fig. 5a), we conclude that the blue and red set of peaks respectively describes centered or lopsided transverse modes. We may partially explain the transition of dominant transverse modes using the diffraction of electron waves. In a deformed cavity, electrons are on average launched at an angle just like in Fig. 5d. These electrons can occupy centered transverse modes (Fig. 5f) only after being considerably diffracted towards the central cavity axis. However, the diffraction angle becomes narrower when the wavelength decreases, so the centered modes become less populated at higher $V_M$. That said, this semiclassical picture neither explains why such centered and lopsided modes should coexist in the cavity nor why the centered mode should ever dominate the deformed cavity as seen in Fig. 5a; the proper theoretical treatment of electronic modes in the open cavity resonator is left for future studies.

In summary, we have observed the formation of electronic modes in an open cavity resonator. Despite the strong coupling to its open sides, the cavity supported well-defined modes which were identified by resonance peaks in the conductance measurements. The regular occurrence of conductance peaks was attributed to longitudinal Fabry-Pérot modes after analyzing the conductance lineshape and resonance energy spectrum. Transverse modes were resolved by introducing a geometric deformation which induced a splitting in the conductance peaks, and the spatial distribution of the modes were investigated from the cavity magnetoconductance. With the aid of tight-binding simulations, we identified two types of resonant modes: the centered type has a conductance maximum at zero magnetic field and a wavefunction lying on the central cavity axis; the lopsided type exhibits a conductance maximum at a finite magnetic field and wavefunction predominantly occupying one side of the cavity axis. Interestingly, we found a transition in the dominant transverse mode as the cavity energy was modulated, but further study is required to fully explain the observed behaviors. Our observations demonstrate a fundamental result bridging the gap between modern cavity optics and open electronic systems by establishing the electronic modes in open cavities.

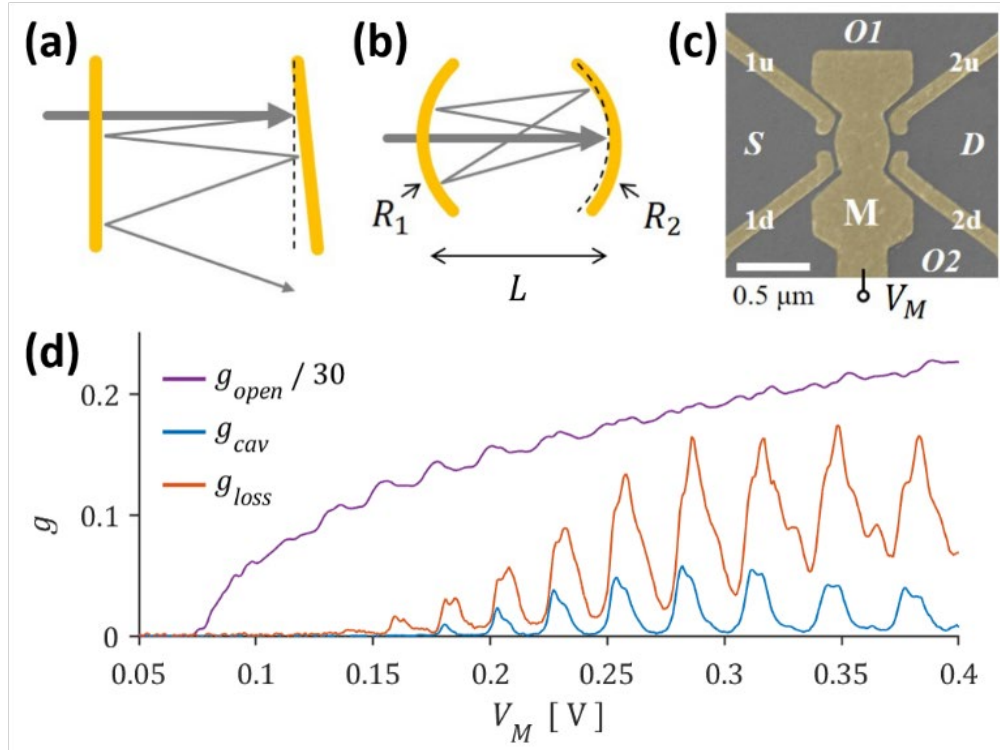

**Fig. 1 Electronic open cavity resonator. a, b** The archetypical Fabry-Pérot Interferometer and the modified Fabry-Pérot resonator. **c** False-colored image of the two-dimensional electron resonator device fabricated on a GaAs/AlGaAs heterostructure. The gates 1u-1d (2u-2d) define the left (right) mirrors of the cavity, and the gate M controls the electron wavelength. The QPCs formed at the center of each mirror was used to tunnel electrons into and out of the cavity. The conductance through the cavity device measured with respect to various lead reservoirs: S, D, O1, and O2. All presented conductances are normalized in units of $2e^2/h$. **d** Various conductances of the cavity measured as a variable of the modulate gate voltage $V_M$. The conductance from O1 to O2, $g_{open}$, shows that the cavity is open in its transverse direction. Despite openness of the cavity, the conductance $g_{cav}$ measured from S to D shows clear signs of resonant transport. The cavity loss, $g_{loss}$, measured from S to O1 and O2, shows concurrent conductance peaks, indicating the diffraction loss of well-defined cavity modes.

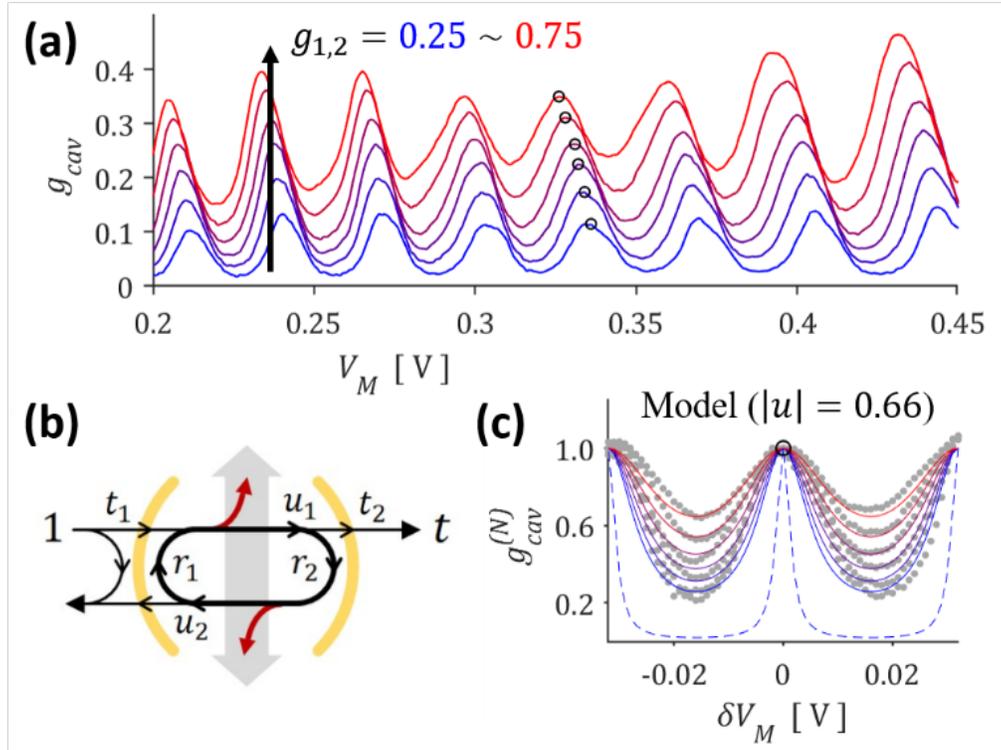

**Fig. 2 Longitudinal Fabry-Pérot modes. a** The cavity conductance $g_{cav}$ measured for various mirror QPC conductances: from $g_{1(2)} = 0.25$ in blue to $g_{1(2)} = 0.75$ in red. As $g_{1(2)}$ decreased, the conductance peak heights did not increase as would be expected from a lossless Fabry-Pérot interferometer. **b** The schematic model of a cavity resonator in the Landauer-Büttiker formalism, where diffraction losses enter as $|u_i| < 1$. **c** The circled peaks (**a**) replotted as grey points after normalization and translation in the $V_M$ axis. The conductance lineshape predicted by a lossy Fabry-Pérot model ($|u_i| = 0.66$) has been plotted over the normalized data for $|t_i|^2 = 0.25$ in blue to $|t_i|^2 = 0.75$ in red. For comparison, the lossless case ($|u_i| = 1$) for $|t_i|^2 = 0.25$ has been plotted as a blue dashed line. The diffractive behavior of electronic waves can be inferred from the agreement between experimental data and the lossy Fabry-Pérot model.

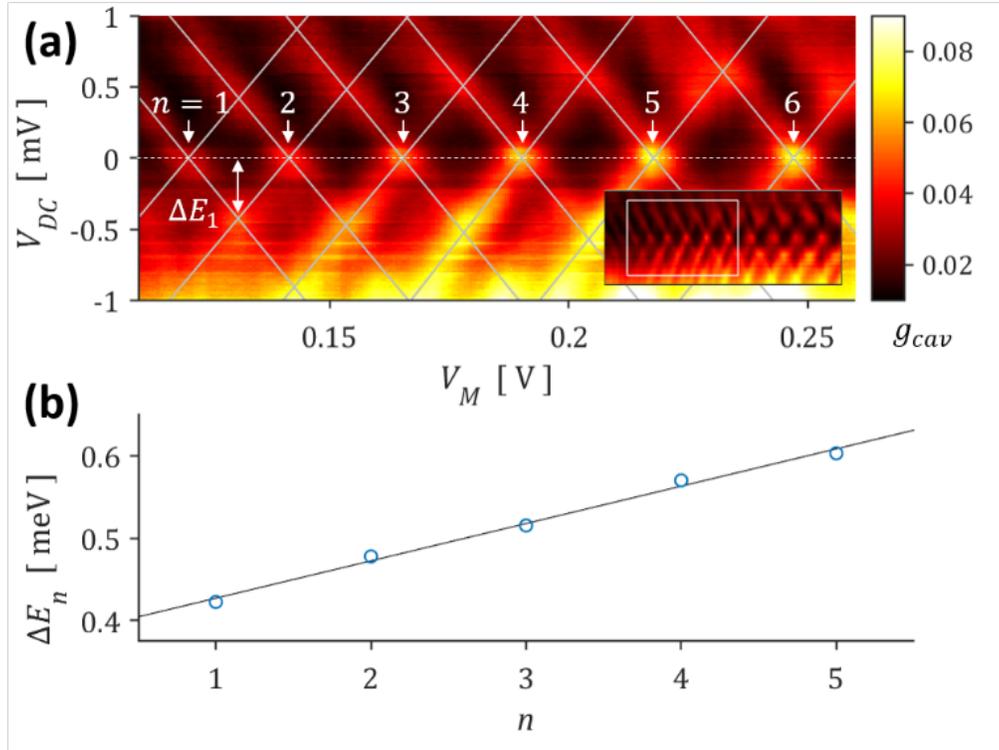

**Fig. 3 Resonance energy spectrum. a** Source-drain bias spectroscopy obtained by applying $V_{DC}$ to reservoir S while measuring $g_{cav}$. The reservoirs O1 and O2 were floated so that the voltage drops only occurred across the mirror QPCs. The energy level spacing between the resonant cavity modes, labeled $n$, can be inferred from $\Delta E_n$. **b** From the full data (inset), peaks without additional fine structures have been analyzed for their level spacings. The linearity $\Delta E_n \sim n$ is from the free electrons resonating in the cavity. The linear fit give $L = 498$ nm, signifying that the resonances correspond to Fabry-Pérot modes forming in the longitudinal direction of the cavity.

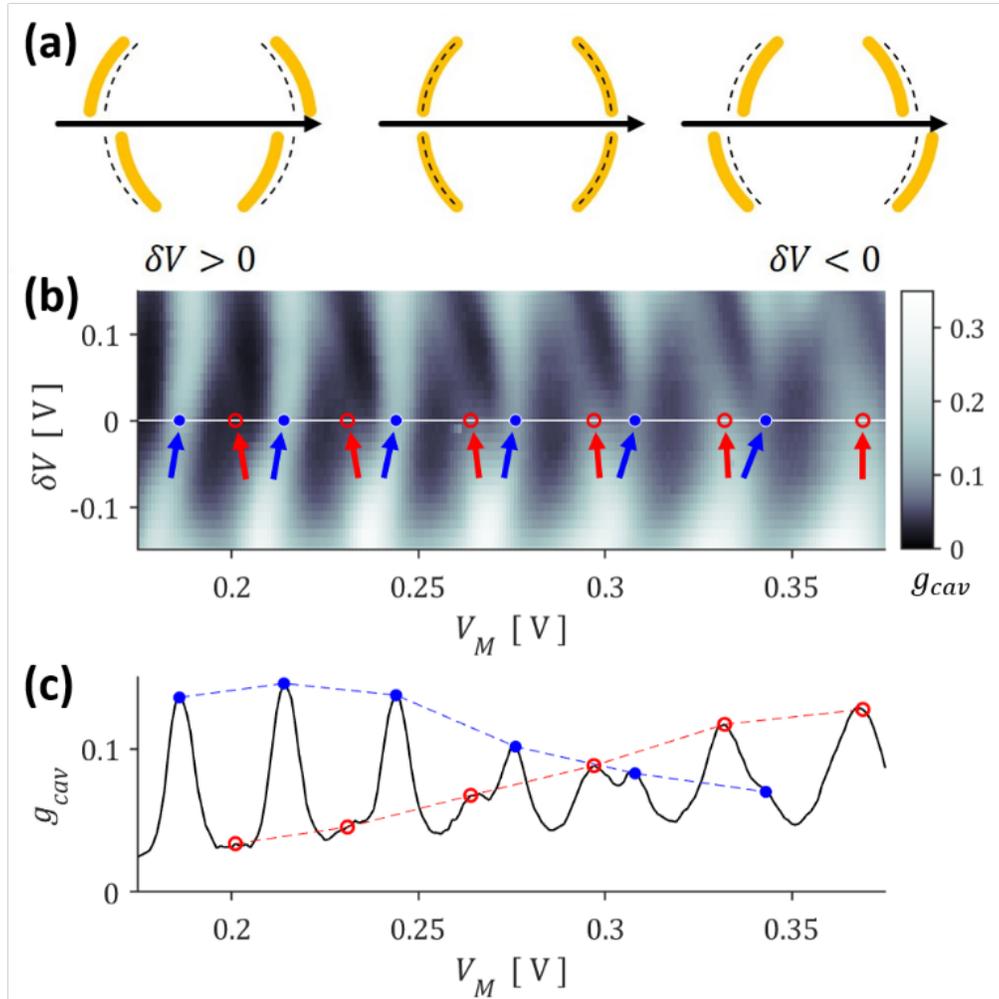

**Fig. 4 Transverse mode detuning. a** The cavity deformation from $\delta V = V_{1(2)u} - V_{1(2)d}$ breaks the reflection symmetry across the central cavity axis. **b** Transverse mode detuning observed by measuring $g_{cav}$ as a function of $\delta V$. Two transverse modes can be identified: one marked with red empty circles and the other marked with blue filled circles. **c** As the electron wavelength is changed, the dominant transverse mode undergoes a transition.

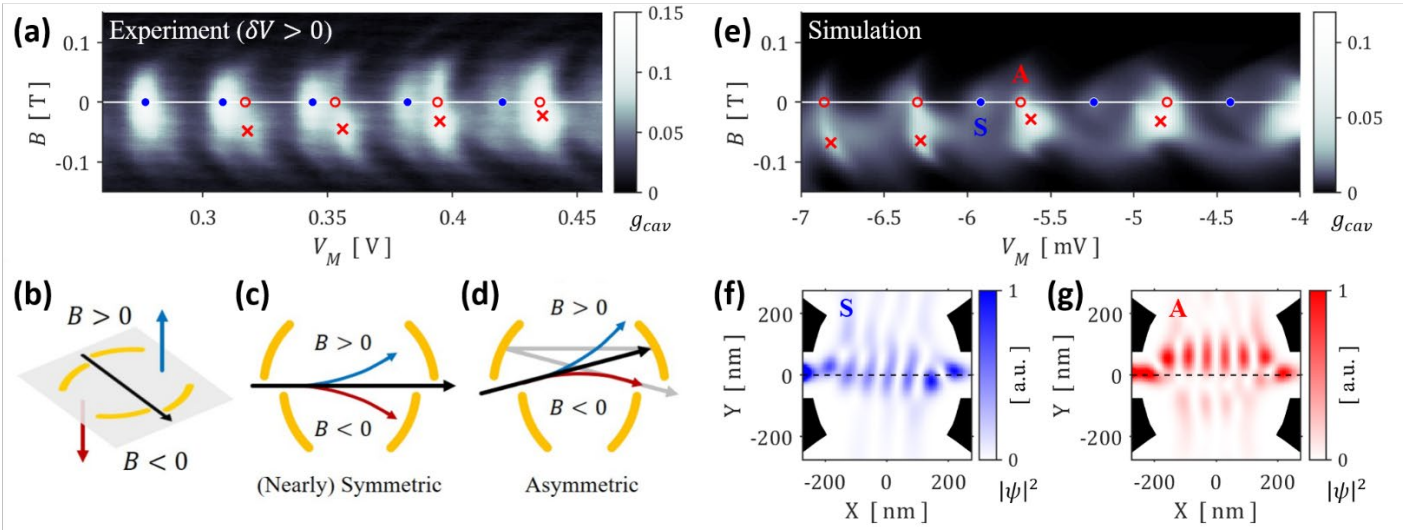

**Fig. 5 Cavity mode distribution. a** The magnetoconductance of the cavity measured as a Lorentz force was applied to the cavity electrons from an out-of-field magnetic field $B$ (**b**). **c** If the electrons occupy the center of the cavity, their response is nearly symmetric with respect to the sign of $B$. Conductance peaks showing this behavior are marked with blue filled circles (**a**). **d** However, an electronic distribution lopsided in the transverse direction distinguishes the sign of $B$. In the illustrated example, $\delta V > 0$, a positive $B$ decreases $g_{cav}$ and a negative $B$ increases $g_{cav}$. Conductance peaks showing this behavior are marked with red empty circles (**a**); red crosses indicate where the conductance peak is maximal. **e** The magnetoconductance of a simulated cavity device shows similar characteristics to the experiment, and the peaks have been marked in the same manner. **f**, **g** The wavefunction density of conductance eigenchannels was obtained for the points marked with S and A (**e**). The black regions indicate the mirror gate positions. The calculated mode geometry agrees with the semiclassical analysis (**c**, **d**).

# Supplementary Information for
# "Observation of Electronic Modes in Open Cavity Resonator"

Hwanchul Jung[1*], Dongsung T. Park[2*], Seokyeong Lee[2], Uhjin Kim[3], Chanuk Yang[3], Jehyun Kim[4], V. Umansky[5], Dohun Kim[4], H.-S. Sim[2], Yunchul Chung[1†], Hyoungsoon Choi[2‡], Hyung Kook Choi[3§]

[1]Department of Physics, Pusan National University, Busan 46241, Republic of Korea.
[2] Department of Physics, KAIST, Deajeon 34141, Republic of Korea.
[3] Department of Physics, Research Institute of Physics and Chemistry, Jeonbuk National University, Jeonju 54896, Republic of Korea.
[4] Department of Physics and Astronomy, Seoul National University, Seoul 08826, Korea.
[5] Department of Condensed Matter Physics, Weizmann Institute of Science, Rehovot 76100, Israel.

## Supplementary Text

### 1. Landauer-Büttiker Model of a Cavity Resonator

In similar spirit to optical systems[1], the open cavity resonator can be treated in the Landauer-Büttiker formalism via scattering matrices. An open cavity has three types of measurement leads: two corresponding to the electron reservoirs beyond the cavity mirrors, and one corresponding to the open sides. Three S-matrices are defined in this picture: $S^{(1,2)}$ describing the tunneling across the cavity mirror (indexed $i = 1,2$) and $S^{(0)}$ describing the intra-cavity reflections and diffractive loss to the open sides. See Fig. S1 for the overview of the model. Note that the variables in the following discussion is either in matrix or block matrix form.

The S-matrix of the cavity mirror $i$ is written as

$$S^{(i)} : \begin{pmatrix} a_i \\ B_i \end{pmatrix} \rightarrow \begin{pmatrix} b_i \\ A_i \end{pmatrix} = \begin{pmatrix} r'_i & t'_i \\ t_i & r_i \end{pmatrix} \begin{pmatrix} a_i \\ B_i \end{pmatrix} \qquad \text{Eq. S1}$$

where $a_i$ ($B_i$) are the amplitudes of modes heading towards the cavity mirror from the lead (cavity) side, $b_i$ ($A_i$) those heading away from the cavity mirror towards the lead (cavity) side, and $r_i$ or $t_i$ ($r'_i$ or $t'_i$) the reflection or tunneling amplitudes of the mirrors heading into (out of) the cavity. By using $a_{12} = (a_1, a_2)^t$ and $b_{12} = (b_1, b_2)^t$, we may rewrite this in the succinct form

$$S^{(12)} : \begin{pmatrix} B \\ a_{12} \end{pmatrix} \rightarrow \begin{pmatrix} A \\ b_{12} \end{pmatrix} = \begin{pmatrix} r & t \\ t & r' \end{pmatrix} \begin{pmatrix} B \\ a_{12} \end{pmatrix} \qquad \text{Eq. S2}$$

which describes all the scatterings across the cavity mirrors. This notation makes it simpler to work with the intra-cavity scattering and cavity-side coupling, given by

$$S^{(0)} : \begin{pmatrix} a_0 \\ A \end{pmatrix} \rightarrow \begin{pmatrix} b_0 \\ B \end{pmatrix} = \begin{pmatrix} v & l \\ g & u \end{pmatrix} \begin{pmatrix} a_0 \\ A \end{pmatrix} \qquad \text{Eq. S3}$$

where $a_0$ ($b_0$) is the amplitude of modes from the open cavity sides heading into (out of) the cavity, $g$ and $v$ the scattering amplitude from the incoming side modes to into the cavity or back to the sides, $l$ the amplitude of diffraction loss from the cavity to the open reservoirs, and $u$ the transmission amplitude from one side of the cavity mirror to the other. A few small notes. Both sides of the open cavity are incorporated in $a_0$ and $b_0$: the upper/lower parts of the open sides can be resolved by splitting the basis, e.g. $a_0 \rightarrow \left(a_0^{(\text{up})}, a_0^{(\text{down})}\right)^t$, but we have not done so here as it does not add to the discussion. Also, all these matrices are unitary by construction, and the diffraction 'loss' term $l$ is lossy in the sense that electrons exit the cavity through the sides. Specifically, we note that $\|u\|^2 + \|l\|^2 = \dim(u)$ where $\|u\|^2 = \text{Tr}(u^\dagger u)$. In a lossless cavity, $\|l\|^2 = 0$ and all the eigenvalues of $u$ are in the simple form $u \sim \exp(i\theta)$ as in the usual Fabry-Perot model. The total cavity S-matrix we wish to find can be algebraically solved for[2]:

$$S: \begin{pmatrix} a_0 \\ a_{12} \end{pmatrix} \rightarrow \begin{pmatrix} b_0 \\ b_{12} \end{pmatrix} = \begin{pmatrix} v + lr[I - ur]^{-1}g & l[I - ru]^{-1}t \\ t'[I - ur]^{-1}g & r' + t'[I - ur]^{-1}ut \end{pmatrix} \begin{pmatrix} a_0 \\ a_{12} \end{pmatrix} \qquad \text{Eq. S4}$$

where we see that $[I - ur]^{-1}$ corresponds to the enhancement factor, as described in optical systems, which give the Fabry-Perot spectrum its signature lineshape[3]. Note that $[I - ur]^{-1}u = u[I - ru]^{-1}$ and $r[I - ur]^{-1} = [I - ru]^{-1}u$ for invertible[4] $r$ and $u$.

For cavity resonances, we are usually interested in $a_{12} \rightarrow b_{12}$ and $a_{1(2)} \rightarrow b_{2(1)}$ in particular. Here, we assume that there is not additional scattering within the cavity, i.e. a rightwards propagating wave does not move leftwards before hitting the mirror. This assumption is mathematically expressed as $u = \text{diag}(u_1, u_2)$, and the full form of $S: (a_1 \quad a_2 \quad a_0)^t \rightarrow (b_1 \quad b_2 \quad b_0)^t$ is written out below for reference:

$$S = \begin{pmatrix} r'_1 + t'_1 u_2 r_2 \dfrac{1}{I - u_1 r_1 u_2 r_2} u_1 t_1 & t'_1 \dfrac{1}{I - u_2 r_2 u_1 r_1} u_2 t_2 & t'_1 \dfrac{1}{I - u_2 r_2 u_1 r_1}(g_2 + u_2 r_2 g_1) \\ t'_2 \dfrac{1}{I - u_1 r_1 u_2 r_2} u_1 t_1 & r'_2 + t'_2 u_1 r_1 \dfrac{1}{I - u_2 r_2 u_1 r_1} u_2 t_2 & t'_2 \dfrac{1}{I - u_1 r_1 u_2 r_2}(u_1 r_1 g_2 + g_1) \\ (l_1 + l_2 r_2 u_1) \dfrac{1}{I - r_1 u_2 r_2 u_1} t_1 & (l_1 r_1 u_2 + l_2) \dfrac{1}{I - r_2 u_1 r_1 u_2} t_2 & V \end{pmatrix}$$

where

$$V = v + l_1 r_1 [I - u_2 r_2 u_1 r_1]^{-1}(g_2 + u_2 r_2 g_1) + l_2 r_2 [I - u_1 r_1 u_2 r_2]^{-1}(u_1 r_1 g_2 + g_1)$$
$$= v + (l_1 r_1 + l_2 r_2 u_1 r_1)[I - u_2 r_2 u_1 r_1]^{-1} g_2 + (l_1 r_1 u_2 r_2 + l_2 r_2)[I - u_1 r_1 u_2 r_2]^{-1} g_1.$$

Note that $[I - ur]^{-1}$, or nearly equivalently $[I - ru]^{-1}$, is present in almost every term, in agreement with optical systems where all transmission from a Fabry-Perot cavity is proportional to the enhancement factor[3]. From the matrix, we can read off $t_{12} = \partial b_1/\partial a_2 = S_{12}$ as

$$t_{12} = t'_1 \dfrac{1}{I - u_2 r_2 u_1 r_1} u_2 t_2. \qquad \text{Eq. S5}$$

and $t_{21}$ trivially given by switching $(1 \leftrightarrow 2)$. As a special case of $S$ being proportional to the enhancement factor, we see that the diffraction losses $S_{3i}$ have terms $[I - r_1 u_2 r_2 u_1]^{-1}$ and $[I - r_2 u_1 r_1 u_2]^{-1}$. Specifically, note that $t_{12} = t'_1[I - u_2 r_2 u_1 r_1]^{-1} u_2 t_2 = t'_1 u_2 [I - r_2 u_1 r_1 u_2]^{-1} t_2$ contains the same term, implying that diffraction loss current shares the same lineshape as the cavity transmission through the mirrors.

## 2. Device Simulation using KWANT

The open cavity device was simulated using the tight-binding numerical package KWANT[5]. A tight-binding Hamiltonian for a spinless two-dimensional electron gas (2DEG) is typically given in the form

$$H = \sum_{ij} U(i,j)|ij\rangle\langle ij| - \sum_{\langle ij,kl \rangle} t(i,j;k,l)|ij\rangle\langle kl| \qquad \text{Eq. S5}$$

where $(i,j)$ is the site index, $\langle ij, kl \rangle$ the notation for $(i,j)$ and $(k,l)$ being neighbor sites, $t(i,j;k,l)$ the hopping term from site $(i,j)$ to site $(k,l)$, and $U(i,j)$ the onsite term. Note that $t(i,j;k,l) = t(k,l;i,j)^*$ for $H$ to be Hermitian. The transport properties of the system are calculated by computing the scattering matrix[2] between semi-infinite leads attached to the boundaries of $H$.

As shown in Fig. S3a, a scattering region of 500 nm × 500 nm was simulated, spanned by a square lattice of spinless electron sites with lattice spacing $a = 5$ nm, and the usual effective mass of a two-dimensional electron in GaAs/AlGaAs heterostructure was used, i.e. $m^* = 0.067 \times m_e$ where $m_e$ is the bare electron mass. The hopping parameter was assigned accordingly: $|t| = \hbar/2m^*a^2$. Using the bare electron charge $-e$, a magnetic field was imposed by the Peierls substitution using a gauge that respects all translation symmetries of the semi-infinite leads[6]. As for the gates, the lithographic geometry used in device fabrication were copied into the simulation. Assuming that the wafer surface lies 50 nm above the 2DEG, the electrostatic gate potentials $\phi$ were calculated in the pinned-potential boundary condition[7]. The potentials were incorporated in the onsite term: $U(i,j) = \phi(x_i, y_j) + 4|t|$ where the $4|t|$ term merely repositions the 2D band minimum at zero energy. The conductances were calculated at a Fermi energy of 8.2 meV as is expected from the electron density in our devices. During calculation, the spatial and energetic dimensions were respectively normalized in units of nm and eV.

Figure S3b shows the mirror QPC conductances as a function of the modulation gate voltage and the mirror gate voltage where a detuning voltage of 15 mV is already in place. Figure S3c shows the potential landscape of the simulated device corresponding to Fig. 5f. Figure S3d shows the cavity conductance for various QPC conductances as a function of the modulation gate voltage. The red line corresponds to the conditions at which the magnetoconductance Fig. 5e was calculated.

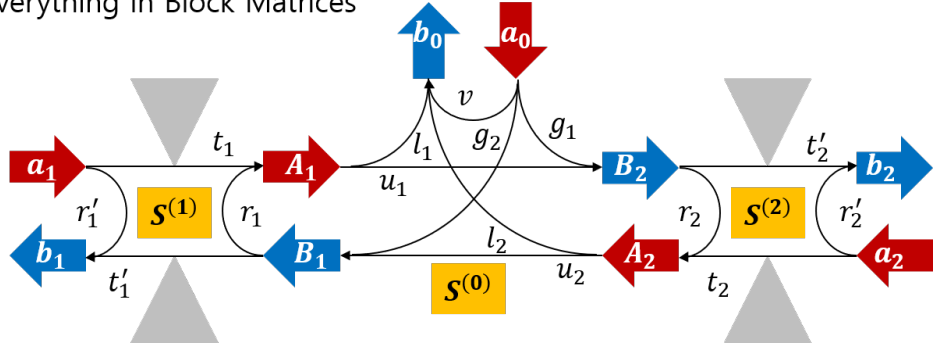

**Fig. S1 Open cavity resonator model.** Incoming and outgoing modes have been indicated with $a_i$, $b_i$, $A_i$, and $B_i$. The measurement lead modes have been written in lowercase, and the cavity resonator modes have been written in uppercase. The cavity mirrors, i.e. QPCs, are described by the scattering matrices $S^{(1,2)}$, whereas the intra-cavity scattering and the coupling to the open sides are described within $S^{(0)}$. The scattering matrix between measurement leads can be algebraically found in the usual manner by cancelling the appearances of $A_i$ and $B_i$. See supplementary text for the details of derivation.

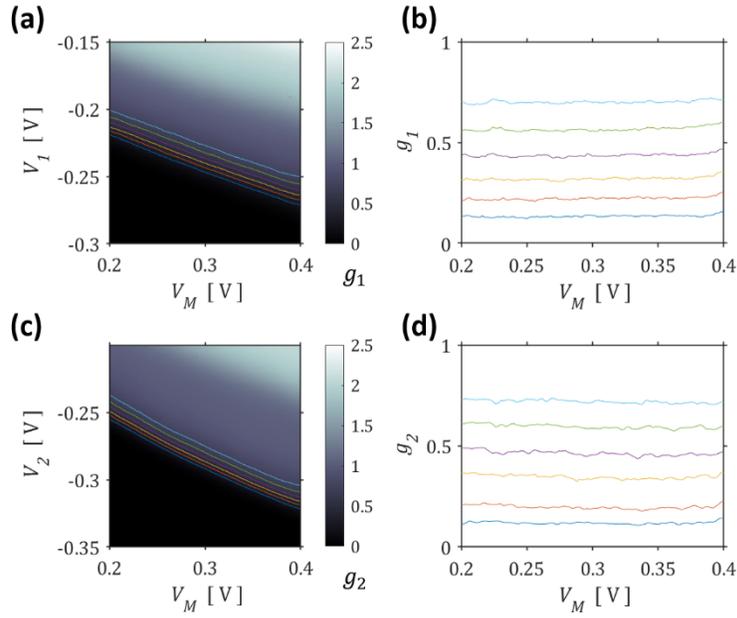

**Fig. S2 QPC tuning. a** The QPC conductance $g_1$ for the mirror 1u-1d plotted as a function of the modulation gate voltage and the mirror gate voltages $V_1 = V_{1u} = V_{1d}$. For $g_1 = 0.25 \sim 0.75$, the voltages $V_1$ for which the QPC conductance is constant has been found using interpolation. These lines have been traced on the plot. The conductance $g_1$ measured while tracing these lines have been replotted (**b**). The conductances indeed maintain a relatively constant value. **c, d** Similar analysis for the mirror 2u-2d.

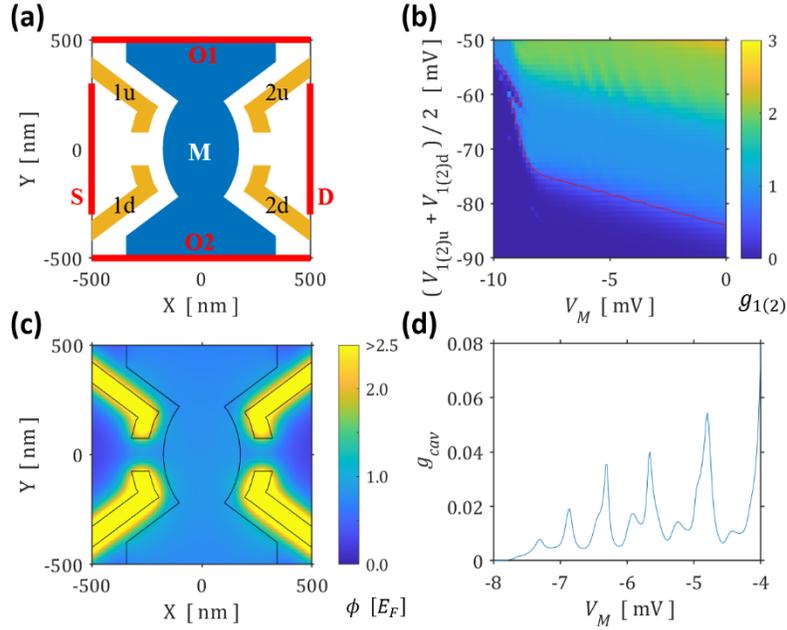

**Fig. S3 KWANT simulation parameters. a** The schematic illustration of the simulated device. The mirror gates have been shaded yellow, and the modulation gate has been shaded blue. The thick red borders correspond to the leads. **b** The conductance of the mirror QPCs as a function of the modulation gate voltage $V_M$ and the mean mirror gate voltages $(V_{1(2)u} + V_{1(2)d})/2$. Note that the mirror gates have already been detuned to $\delta V = V_{1(2)u} - V_{1(2)d} = 15$ mV. The red line traces where the QPC conductance is fixed at $g_{1,2} = 0.5$. **c** The potential $\phi$ incurred by the gates at the conditions for Fig. 5f. The edges of the gates have been delineated with black lines, and the colorbar has been given in units of the Fermi energy $E_F = 8.2$ meV. **d** $g_{cav}$ calculated along the red line (**b**). This plot corresponds to the $B = 0$ data in Fig. 5e. See supplementary text for detailed model and simulation parameters.